\documentstyle[epsf]{MN}

\def \mpc       {{\rm\ Mpc}}

\def \ie        {\hbox{\it i.e.}}
\def \eg        {\hbox{\it e.g.}}

\def \etal      {{\it et al.\ }}

\def \Ho        {{\rm\ H_{o}}}
\def \kmsmpc    {{\rm\ km\ s^{-1}\ Mpc^{-1}}}
\def \kev       {{\rm\ keV}}
\def \msol      {{\rm M}_\odot}
\def \h         {\hbox{$\, h$} }
\def \hinv      {\hbox{$\, h^{-1}$} }

\def \se        {\!=\!}

\def \ssim      {\! \sim \!}

\def \sequiv    {\! \equiv \!}

\def\\{\hfil\break}
\def\spose#1{\hbox to 0pt{#1\hss}}
\def\lta{\mathrel{\spose{\lower 3pt\hbox{$\mathchar"218$}}
     \raise 2.0pt\hbox{$\mathchar"13C$}}}
\def\gta{\mathrel{\spose{\lower 3pt\hbox{$\mathchar"218$}}
     \raise 2.0pt\hbox{$\mathchar"13E$}}}

\newcount\itemno
\def \ino         { \the\itemno\global\advance\itemno by 1 }
\def\apj{ApJ}
\def\aj{AJ}

\def\apjs{ApJS}
\def\aa{A\&A}
\def\mnras{MNRAS}
\def\araa{ARA\&A}
\def\nature{Nature}

\def \xray {\hbox{X--ray} }
\def \rfiveh {\hbox{$r_{500}$}}
\def \Mfiveh {\hbox{$M_{500}$}}
\def \rdeltac {\hbox{$r_{\delta_c}$}}
\def \rvir {\hbox{$r_{vir}$}}
\def \rhocrit {\hbox{$\rho_c$}}
\def \fbargas {\hbox{${\bar f}_{gas}$}}

\def \einstein {\hbox{\it Einstein} }

\title [ The Intracluster Gas Fraction in X--ray Clusters ] 
{ The Intracluster Gas Fraction in X--ray Clusters : ~~~~ 
Constraints on the Clustered Mass Density }
 
\author [ August E. Evrard ] { August E. Evrard$^{1,2}$ \\ 
$^1$ Physics Department, University of Michigan, Ann Arbor, MI 48109-1120 USA \\
$^2$ Institut d'Astrophysique, 98bis Boulevard Arago, F75014 Paris, France \\ 
E-mail address : evrard@umich.edu }

\date{Submitted for publication in MNRAS}

\begin{document}

\maketitle

\begin{abstract}

The mean intracluster gas fraction 
of \xray clusters within their hydrostatic regions  
is derived from recent observational compilations
of David, Jones \& Forman and White \& Fabian.  
At radii encompassing a mean density 500 times 
the critical value, the individual 
sample bi-weight means are moderately ($2.4 \sigma$) discrepant; 
revising binding masses with a virial relation 
calibrated by numerical simulations removes the discrepancy and 
results in a combined sample mean and standard error 
$\fbargas(\rfiveh) = (0.060 \pm 0.003) \h^{-3/2}$. 
For hierarchical clustering models with  
an extreme physical assumption to maximize cluster gas content, 
this value constrains the universal ratio of total, clustered 
to baryonic mass 
$\Omega_m/\Omega_b \le 23.1 \h^{3/2}$; 
combining with the primordial nucleosynthesis upper limit on 
$\Omega_b$ results in $\Omega_m \h^{1/2} < 0.60$.  A less 
conservative, physically plausible approach based on low 
D/H inferences from quasar absorption spectra and 
accounting for baryons within cluster galaxies yields an estimate 
$$
\Omega_m \h^{2/3} = 0.28 \pm 0.07
$$
with sources of systematic error involved in the derivation 
providing approximately $35\%$ uncertainty.  Additional effects 
which could provide consistency with the Einstein--deSitter case 
$\Omega_m \se 1$ are presented, and
their observable signatures discussed.  

\end{abstract}


\section{Introduction}

Clusters of galaxies provide a number of interesting cosmological 
diagnostics.  In particular, the relative amount of baryons and dark matter 
within their hydrostatic regions provides a measure of the cosmic mix  
of these components, \ie, a measure of the ratio of density parameters 
$\Omega_m/\Omega_b$ in the Friedmann--Lema\^{\i}tre world 
model\footnote{In this paper, $\Omega_m$ refers to
the contribution of all clustered matter (including baryons) 
to the stress--energy density.}

Employing this 
measure in practice requires accurate observational data 
along with estimates of possible systematic biases.  
Likely sources of bias include systematic errors in component mass estimates 
and deviation 
of the local cluster ratio of baryonic--to--total mass arising, 
for example, from the 
different dynamical histories of the two components.   Within the context 
of hierarchical clustering scenarios, these effects
have been calibrated by numerical simulations of cluster formation.  
The results, discussed in detail below, indicate that 
the magnitude of these effects are small 
--- a few tens of percent or less --- in 
observationally accessible regions of clusters.  If our current
description of cluster formation dynamics is physically accurate, 
then determination of the cosmic baryon fraction from \xray cluster 
data is straightforward.  

The baryonic component of the richest clusters is dominated by the 
\xray emitting 
intracluster gas rather than the mass associated with the optical light of 
the cluster galaxies (Forman \& Jones 1982; Sarazin 1986; Mushotzky 1994).  
Taking Coma as an example, White \etal (1993) 
estimate the ratio of gas to galaxy mass to be 
$M_{gas}/M_{gal} = (5.5 \pm 1.5) \h^{-3/2}$ 
(with $h \! = {\rm H}_0/100 \kmsmpc$).  This number is typical of the 
values seen in larger samples for clusters comparable in temperature
to Coma (David \etal 1990; Arnaud \etal 1992).
The data also indicate a dependence of the ratio $M_{gas}/M_{gal}$ 
on cluster temperature, with poorer clusters and groups possessing 
less gas per massive galaxy than their larger counterparts.  

The gas mass thus provides not only a formal lower limit to the total
baryon cluster content, but a fairly accurate estimate of the baryon mass in 
rich clusters if the dark matter is assumed non--baryonic. 
Recent compilations of \xray cluster data by White \& Fabian (1995) 
and David, Jones \& Forman (1995) provide gas and total 
mass estimates for samples of moderate size.  
In this paper, I use these data, along with guidance from numerical 
simulations, to estimate the sample mean gas 
mass fraction $\fbargas$ within a characteristic radius 
defining the hydrostatic boundary of clusters.  
Motivations for the specific approach adopted here are provided in \S2
and the observational data are analysed in \S3.   
Implied constraints on the universal baryon fraction and the value of
$\Omega_m$ are provided in \S4, including a thorough discussion of 
systematic errors.  

\section{Physical Expectations}

The value of any individual cluster's baryon-to-total 
mass inferred from observations will, in general, differ 
from the cosmic value.  
The difference can be partly intrinsic --- reflecting a true baryon
enhancement or deficit within the cluster --- and partly due to errors
in estimates of the component masses.  

To the extent that the physical processes responsible for a cluster's  
structure are independent of its total mass, any intrinsic bias in 
the baryon fraction can, to first approximation, be expressed as 
a function of a scaled radial variable, equivalent to the 
local density contrast.  Introducing notation used below, 
$\delta_c$ denotes the mean 
interior density contrast with respect to the critical value 
$\delta_c \sequiv \rho(<r)/\rhocrit$ with $\rhocrit \equiv 3
\Ho^2/8\pi$G.  Similarly, $\rdeltac$ is the radius at which a 
density contrast $\delta_c$ is attained.  

\subsection{Intrinsic component bias}

Because the horizon mass scale at the baryogenesis epoch is 
many orders of magnitude smaller than the typical 
cluster mass (\eg, Kolb \& Turner 1990), there is no causal mechanism 
which can generate primordial fluctuations in the
baryon--to--total mass ratio on cluster scales.  
This argument holds as long as inflation precedes baryogenesis.  
Any intrinsic bias must therefore be set by dynamical processes operating
differentially on the baryonic and dark matter. 

Gravity alone is not a powerful segregating mechanism.   
In the simplest example, consider clusters forming from 
spherically symmetric, scale--free initial
density perturbations.  Self--similar solutions to the dynamical 
equations (Fillmore \& Goldreich 1984; Bertschinger 1985; 
Chi\`eze, Teyssier \& Alimi 1996) 
exhibit two characteristic regions ---  
a nearly hydrostatic, inner body surrounded by an outer, infalling envelope 
which merges seamlessly into the Hubble flow at large radii.  
The flow changes discontinuously at the boundary, implying the 
development of a shock for the collisional baryons or a caustic surface 
for the collisionless dark matter (or galaxies).  The position and velocity 
of the shock for a $\gamma \se 5/3$ ideal gas is very close to that of
the outermost caustic surface 
for the dark matter; together they define a unique radius (commonly 
referred to as the {\it virial\/} radius \rvir) within which the 
mean enclosed density is $\sim 100$ times the background value.  
Since all matter outside the virial surface is infalling with the  
cosmic mix of components, then, by continuity, the baryon mass
fraction measured at the virial radius {\it must\/} be unbiased 
$f_b (\rvir) \equiv \Omega_b/\Omega_m$.

The realistic case differs from the spherical one in several
respects.  Clusters forming in a fully three--dimensional, 
hierarchical fashion experience lumpy, asymmetric accretion 
directed by connecting filaments 
(West, Dekel \& Oemler 1987; Frenk \etal 1990;  
Evrard 1990a,b; Kang \etal 1994; Navarro, Frenk \& White 1995; 
Tormen 1996).  
In addition, the ability of the intracluster gas
to lose entropy via radiative cooling or increase it
via feedback from galactic winds calls into question 
lessons learned assuming gravitationally induced shocks are 
the only entropy changing mechanism.  

These issues led White \etal (1993) to explore extreme models for
dynamical baryon enhancement.  They examined the evolution of 
infinitely dissipational (pressureless) gas and dark matter using 
both a spherical model based on Bertschinger's (1985) solutions and
three dimensional, gas dynamic simulations.  
Following their notation, define $\Upsilon(\delta_c)$ as the ratio 
of enclosed baryon fraction within radius $\rdeltac$ to the cosmic value
\begin{equation}
f_b (\rdeltac) \ \equiv \ \Upsilon(\delta_c) \ \frac{\Omega_b}{\Omega_m} .
\label{fb_rdeltac}
\end{equation}
For the extreme case of a zero temperature gas, 
the White \etal three dimensional simulations show baryon enhancements
smaller than the spherical case.  For example, at $\delta_c \se 500$, 
the spherical model predicts $\Upsilon(500) \se 1.6$ while the average
of twenty simulations is $\Upsilon(500) \se 1.25$ and 
the maximum simulation value is $1.5$.  Although the simulation
results strictly apply to the case of a standard CDM initial fluctuation 
spectrum, results for other cosmologically reasonable $\Omega_m=1$ power
spectra should not differ substantially, as the sensitivity of 
cluster dynamical histories to spectral shape is fairly mild 
(Lacey \& Cole 1993; Crone, Evrard \& Richstone 1994; 
Navarro, Frenk \& White 1996).  

When a realistic gas equation of state is used, combined N-body and gas 
dynamic simulations generally show the gas to be slightly more
extended than the dark matter (Evrard 1990a; Thomas \& Couchman 1992; 
Cen \& Ostriker 1993; Kang \etal 1994; Metzler \& Evrard 1994; 
Navarro, Frenk \& White 1995; Lubin \etal 1996), 
though the evidence is not universal (Anninos \& Norman 1996).   
Energy transfer between these components during major 
mergers is a likely physical explanation (Navarro \& White 1993; 
Pearce, Thomas \& Couchman 1994), though the origin may be unrelated 
to mergers (Chi\'eze, Teyssier \& Alimi 1996).  
The extended gas structure implies a weakly rising 
baryon fraction with radius 
$f_{gas}(r) \ssim r^\eta$ with $\eta \ssim 0.1-0.2$ near the 
virial radius, and a modest, overall baryon  
diminution $\Upsilon(500) \ssim 0.9$ 
(Frenk \etal 1996).  The mild rise of gas fraction with radius 
appears consistent with the observations discussed below.  

\subsection{Cluster mass estimation}

The bias and variance in the mass estimates inferred from observations
have also been calibrated by numerical experiments of cluster
formation incorporating gravity and gas dynamics.  
A number of independent experiments over the years 
(Evrard 1990a,b;  Tsai, Katz \& Bertschinger 1994; 
Metzler \& Evrard 1994; Navarro, Frenk \& White 1995; Schindler 1996; 
Evrard, Metzler \& Navarro 1996; 
Roetigger, Burns \& Loken 1996) show that, when exercised 
judiciously, accurate binding mass estimates can be made with the 
standard, $\beta$-model approach (Cavaliere \& Fusco-Fumiano 1976).  
``Judicious exercising'' here means avoiding the cores of clusters
where cooling flows and other complications occur (Tsai, Katz \&
Bertschinger 1994), avoiding clusters engaged in obvious 
major mergers (Roetigger \etal 1996), and avoiding extrapolating 
to very large radii where an equilibrium assumption is not 
justifiable.  

From an analysis of gas velocity moments, 
Evrard \etal (1996, hereafter EMN)
propose $\rfiveh$ as a conservative choice for the outer 
hydrostatic boundary of clusters.  Using a sample of 56 cluster 
simulations evolved in different cosmological backgrounds, and with a 
subset including input of mass and energy from galactic winds, they
show that, within this radius, the gas is very close to hydrostatic, 
with a mean, mass weighted, radial Mach number of only a few percent 
(see Tables~3 and 4 of EMN).  Binding mass estimates based on the 
standard, $\beta$--model approach are nearly unbiased and 
have an intrinsic scatter of $\sim 30\%$ at $\rfiveh$.   
This compares well with the $15\%$ {\it rms}\ deviation quoted by 
Schindler (1996) and the tens of percent variations seen in the 
experiments of Roettiger \etal (1996).

In addition, EMN find that the variance in the mass estimates 
can be considerably reduced by eliminating the $\beta$ parameter entirely 
({\it i.e.,\/} ignoring the \xray image) and 
deriving the binding mass directly from the global, emission weighted 
temperature $T_X$.  The resulting scaling relations are consistent 
with virial equilibrium expectations at a fixed density contrast
$T \sim GM/r \sim \delta_c \rhocrit r^2$.  Calibrating the relation with 
the 56 experiments at $\delta_c \se 500$ yields 
\begin{equation}
\rfiveh(T_X) \ = \ (1.24 \pm 0.09) \ \biggl({T_X \over 10 \kev}\biggr)^{1/2}
\hinv \mpc  \ ,
\label{r500}
\end{equation}
\vskip -0.3 truecm
\begin{equation}
\Mfiveh(T_X) \ = \ (1.11 \pm 0.16) \ \biggl({T_X \over 10
\kev}\biggr)^{3/2} \times 10^{15} \hinv \msol \ .
\label{M500}
\end{equation}
For the numerical sample, a standard deviation of only $15\%$ in 
mass estimates results from this relation.  

Gas masses can be recovered with typically higher accuracy 
than the dark matter 
(Metzler \& Evrard 1994; Roettiger \etal 1996), but this is 
based upon the assumption that the gas is smoothly distributed within 
intracluster space, not bound into filamentary or knotty clumps.  There 
is no well defined physical model for such clumpiness, rather the motivation for
such a model comes from a desire to minimize the gas fraction in clusters
while still providing the observed emission measure for bremsstrahlung.  
Since the latter scales as $\rho_{gas}^2$ while the former as $\rho_{gas}$, 
a large, local "clumping factor" $C \equiv <\rho_{gas}^2>^{1/2} / \rho_{gas}$ 
could reduce the amount of gas required to produce a fixed \xray 
emission by a factor $\sim \! C$.  

There is some observational evidence 
arguing against such clumping.  A very high signal--to--noise 
\xray image of the Coma cluster shows no signs of fluctuations in the 
emission apart from that associated with individual galaxies (White, Briel
\& Henry 1993).  In addition, measurements of the Sunyeav--Zel'dovich 
(SZ) decrement in clusters are consistent with expectations based on no 
significant clumping for reasonable values of the Hubble 
constant (Birkenshaw, Hughes \& Arnaud  1991; Herbig \etal 1995;
Carlstrom, Joy \& Grego 1996).  For example, again in Coma, 
Herbig \etal (1995) derive a 
Hubble constant $\Ho \se 71^{+30}_{-25} \kmsmpc$ from OVRO 
observations of the 
SZ effect combined with the \xray model of Hughes (1989) for the 
intracluster gas, which assumes no clumping.  However, given the present
uncertainty in these and other Hubble constant determinations 
(Kennicutt, Freedman \& Mould 1995), there appears room for a 
systematic effect from clumping at the tens of percent level, 
and perhaps higher.  This and other systematic effects are discussed
in \S4.3 below.  

\section{The Mean Cluster Gas Fraction}

In this section, I determine the mean gas fraction 
within $\rfiveh(T_X)$ --- calibrated by the numerical experiments, 
eq'n(\ref{r500}) --- for the observational samples presented by  
White \& Fabian (1995) and David, Jones \& Forman (1995).  

White \& Fabian (1995; hereafter WF) present gas fractions for 19 
clusters derived from archival \einstein IPC observations and 
temperatures compiled from the literature, many from David \etal (1993).  
They quote values for $f_{gas}$ at both a fixed metric radius of 
$0.5 \hinv \mpc$ and the cluster \xray radius $r_X$ which is governed by the 
image quality.  To evaluate gas fractions at $\rfiveh(T_X)$, I use 
a mild, power law extrapolation of the data quoted at $r_X$
\begin{equation}
f_{gas}(\rfiveh(T_X)) \ = \ f_{gas}(r_X) \ 
 \biggl({\rfiveh(T_X) \over r_X}\biggr)^\eta 
\label{fgasExtrap}
\end{equation}
with $\eta \se 0.17$.  This value of $\eta$ is derived from the
numerical simulations of EMN, and is consistent with the rise 
of gas fraction with radius in both the WF and DJF samples.   
The mean values of $\eta$ in the WF data, derived by comparing the 
baryon fractions at $0.5 \hinv \mpc$ and $r_X$, is $0.13$.  This 
is biased somewhat low by the few clusters with short lever arm --- 
removing 3 clusters with $r_X < 0.6 \hinv \mpc$ yields a mean $\eta
\se 0.15$, with all but one value in the range $0.05 - 0.28$.  


\begin{figure}
\vskip -2.4 truecm
\epsfxsize=9.5cm
\epsfysize=9.5cm

\epsfbox{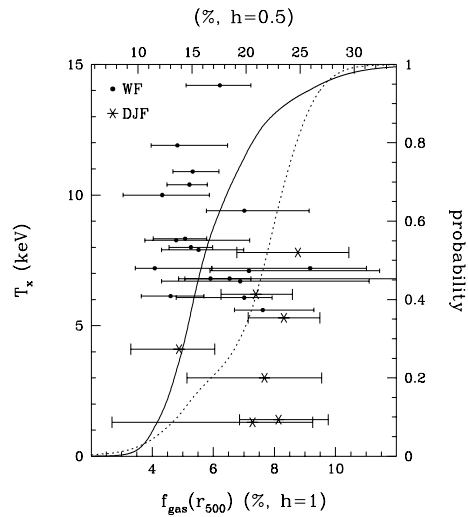}

\caption{The intracluster gas fraction within 
$\rfiveh(T_X)$ from eq'n(\ref{fgasExtrap}) plotted against 
\xray temperature for the observational samples of WF (solid dots) 
and DJF (crosses).  
Errors in the temperature, omitted for clarity, are fractionally 
comparable to those in gas fraction.  
Estimates of the samples' cumulative parent probability distributions,
derived from the normalized sum of the individual cluster asymmetric 
Gaussian contributions, are shown by the solid (WF) and dotted (DJF) 
lines. 
}
\label{fbdat}
\end{figure}

The resultant gas fractions at $\rfiveh(T_X)$ for the WF sample 
are plotted against cluster temperature as filled circles in 
Figure~\ref{fbdat}.  Error bars on the gas fraction are $90\%$ 
confidence limits and include  
contributions from the binding and gas mass errors.  The errors 
quoted in Table~2 of WF include only the gas contribution.  To estimate 
binding mass uncertainty, I assume a fractional error equal to that in 
cluster temperature.  Because the latter is generally asymmetric, 
so also is the binding mass error.  Temperature values and errors are 
taken from the deprojected values in Table~1 of WF, except for two 
cases of very hot clusters --- A2142 and A2163 --- which have
unusually large allowed lower temperature ranges from the deprojection
method.  For these, I use the ``reference'' lower bound on
temperature quoted by WF rather than the deprojected values.  

The temperature errors are comparable in magnitude to those of the
gas; average $1\sigma$ fractional errors are $8.5\%$ for the gas
and $6.6/12.8\%$ for the upper/lower temperature uncertainties.  
Because the gas emissivity in the IPC band is fairly insensitive 
to temperature over most of the range spanned by the WF data, the
derived gas masses are nearly independent of temperature.  Errors 
in gas and total masses can then be assumed uncorrelated, implying the   
fractional square error in $f_{gas}$ is the sum of the squares of 
the fractional errors in the gas and total masses.

David \etal (1995; hereafter DFJ) provide in their Figure~5
the gas fraction as a function of overdensity $\delta_c$ directly.
They present data on a range of systems, 
from elliptical galaxies to clusters; I use only the seven 
groups and clusters with $T_X \! > \! 1 \kev$; namely, 
A539, A262, A2589, A2063, A1795, A85 and A2029.  Errors in the binding
masses for these systems are again taken to be proportional to the
temperature errors listed in Table~1 of DJF.  Gas mass errors are not 
quoted in their Table, for these I assume a fixed $1\sigma$ error of 
$8.5\%$, the mean of
the WF sample.  This is likely to be an overestimate of the actual
typical error, given the improved image quality of the ROSAT PSPC over
the \einstein IPC.  The results are insensitive to the choice 
of assumed error over the range $0-15 \%$.  
The data are plotted as crosses in Figure~\ref{fbdat}.  

The cooler ROSAT sample appears to be more gas rich than its hotter 
IPC counterpart.  Non--parametric statistical tests of location, 
such as the Mann--Whitney test and the 
Wilcoxon Signed-rank test, indicate sample inconsistency 
at better than $99\%$ confidence for the raw data values. 
A simple way to incorporate errors is through 
estimates of the samples' parent probability distributions,  
derived from an unweighted sum 
of the individual cluster asymmetric Gaussian contributions.  
The cumulative version of these are shown for each sample as 
the solid and dotted lines in Figure~\ref{fbdat}.  

The range of likely baryon 
fraction values, as measured by the 5 to 95\% confidence intervals, 
is large, $4 - 10 \h^{-3/2} \%$, and the two data sets provide 
consistent estimates for this range.  However, they differ 
substantially in their 
median values of $5.7 \h^{-3/2}$ (WF) and $7.7 \h^{-3/2}$ (DJF) 
percent, reflecting the fact that 
the distributions are asymmetric and oppositely skewed.  
A likelihood analysis of the data yields most likely values 
of $5.6 \h^{-3/2}$ and $7.3 \h^{-3/2}$, respectively, very 
similar to the median values.  


\begin{figure}
\vskip -2.0truecm

\epsfxsize=8.5cm
\epsfysize=8.5cm
\epsfbox{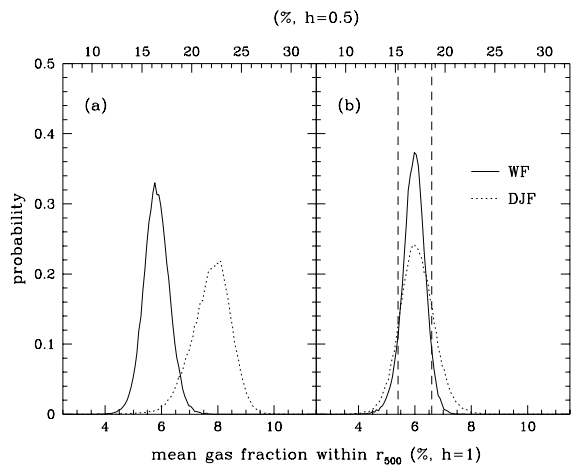}

\caption{Probability distribution function 
of the sample bi-weight mean gas fraction at $\rfiveh$ obtained from 
bootstrap estimation.  The left
panel employs the original binding mass estimates from the literature while 
the right panel replaces the original mass estimates with values 
derived from $T_x$ and the virial scaling relation, eq'n(\ref{r500}). 
The dashed lines show the $95\%$ confidence 
region for $\fbargas(\rfiveh)$ derived from the combined, revised 
samples.
 }
\label{fbMeanDist}
\end{figure}

A more robust assessment of sample location is made by 
employing a bootstrap procedure to estimate the range of 
sample means consistent with each data set.  A bi-weight estimator
of location is used, because of its superior performance for small 
data samples (Beers, Flynn \& Gebhardt 1990), but results using an ordinary 
mean or median are similar.  The procedure creates a large 
number of trial samples with replacement, assuming data values  
distributed in an asymmetric Gaussian fashion.  For each trial, 
the bi--weight mean is estimated and the resulting distribution 
over many trials constructed.  

The results, shown in panel (a) of Figure~\ref{fbMeanDist}, 
are very nearly symmetric and Gaussian, the DJF sample less so 
because of the influence of A2063, an outlier with 
$f_{gas}(\rfiveh) \se 4.9 \h^{-3/2} \%$ (see Figure~\ref{fbdat}).  
The bi--weight means and standard errors of the samples 
--- $5.82 \pm 0.45  \h^{-3/2} \%$ (WF) and 
$7.77 \pm 0.69 \h^{-3/2} \%$ (DJF) --- 
differ at the $2.4 \sigma$ level, where the significance 
quoted is $|x_1-x_2|/\sqrt{\sigma_1^2 + \sigma_2^2}$, 
appropriate for independent Gaussian distributions with 
means $x_1$ and $x_2$ and standard deviations $\sigma_1$ and
$\sigma_2$.  This is supported by the appearance of the 
distributions in Figure~\ref{fbMeanDist}(a).  

The direction of the discrepancy --- lower temperature DJF groups having 
{\it higher} gas fractions than the richer clusters of the WF sample
--- is surprising, because adding the galaxy contribution to the
baryon mass would serve to exaggerate the difference, not correct it.  
Adding the galaxies would imply significantly larger baryon 
fractions in poor groups compared to rich clusters within the virial
radius, an outcome not anticipated in hierarchical clustering
scenarios.  In fact, the opposite is the more likely expectation.  
Feedback from galactic winds drives 
baryons preferentially out of low temperature systems 
(Yahil \& Ostriker 1973; White 1991; Metzler \& Evrard 1994).  
It is possible that the discrepancy arises from sample selection
criteria, but the fact that neither sample is statistically well defined  
makes it difficult to address this issue.  Qualitatively, 
it is difficult to understand why the generally lower quality IPC 
imaging should be biased in favor of lower gas fraction clusters.

A possibility addressed here 
is that modest biases in the binding masses are responsible for 
the discrepancy.  Though both are based on an underlying assumption
of equilibrium, WF and DJF use slightly different approaches for
estimating binding masses.  Only a small 
($\ssim 25\%$) systematic effect is required to bring the two sample 
means into agreement.  Analysis of the numerical sample of EMN 
indicated the virial scaling relation provided a more accurate --- in 
the sense of minimizing variance --- mass estimator than the standard,
$\beta$--model method.  If real galaxy clusters satisfy virial 
equilibrium to the degree established in the simulations, then 
application of eq'n~(\ref{M500}) to the observations should produce 
similarly accurate mass estimates.  This suggests a re-analysis of 
the data, employing revised binding mass estimates derived 
from eq'n~(\ref{M500}) and the measured cluster temperatures $T_x$.  

The result of repeating the 
bootstrap procedure with the revised binding masses is shown in 
Figure~\ref{fbMeanDist}(b).  Appropriate corrections have been made 
to maintain estimates at a density contrast $\delta_c \se 500$.  
The WF data shift by a small amount, 
indicating good agreement between the virial scaling relation and 
the original deprojection mass estimates, and the variance 
decreases slightly.  
The DJF data shift more substantially to lower gas fractions.  
The direction and magnitude of the shift can be 
traced directly to the mean value of $\beta$, which enters linearly in 
the original mass estimates.  The seven groups and clusters of the DJF 
sample have $\bar{\beta} \se 0.63$, about $25\%$ lower than the ``magic'' 
value of $0.79$ inferred from inserting the radial scaling, 
eq'n~(\ref{r500}), into the $\beta$--model estimator (see EMN).   

The revised distributions provide consistent estimates of the mean cluster 
gas fraction between the two samples.  
Combining the data sets results in a highly statistically accurate 
estimate of the bi--weight mean gas fraction within $\rfiveh$ 
for \xray clusters
\begin{equation}
\fbargas(\rfiveh) \ = \ (0.060 \pm 0.003) \ \h^{-3/2} 
\label{fbargas}
\end{equation}
where the quoted error is $1\sigma$.  For comparison, 
the same limits derived from combining the original sample data 
is $0.063 \pm 0.004$.  In aligning the two samples, the revised 
mass estimates provide a small reduction in the uncertainty while 
not significantly affecting the location of the mean.  The analysis which
follows employs the revised gas fraction value, 
but clearly similar numbers result if the original data 
are employed.  

This analysis of the population mean should not be misinterpreted as
presenting {\it the} value for the gas fraction within {\rfiveh} for 
{\it all} clusters.  Simulations indicate intrinsic variations in 
the gas fraction at about the $15\%$ level arise naturally from the 
different dynamical histories/states of a coeval population (White
\etal 1993; Cen \& Ostriker 1993; Kang \etal 1994).   
Galactic feedback can cause gas loss by subsonic winds after cluster
formation or by hindering collapse itself through early preheating. 
Though nearly all the 26 clusters in Figure~\ref{fbdat} have 90\%
confidence limits which overlap the limits in eq'n(\ref{fbargas}), 
there are clusters which have significantly less gas.  An example is 
A576, which has a gas fraction lower by about a factor 2 than the 
mean (Mohr \etal 1996).  Compact groups are extreme in this regard
(Ponman \etal 1996; Pildis, Evrard \& Bregman 1996).  
Loewenstein \& Mushotzky (1996) present two cool clusters 
with different baryon fractions within their virial regions.  
Intrinsic variations in cluster gas content are thus both expected and
observed, with the caveat that large amplitude variations go in only one 
direction, that of reducing gas content.  There is currently no empirical
evidence or theoretical justification supporting large baryon 
enhancements near the virial radius in clusters.  

\section{Implications for $\Omega_m$}

The arguments presented in \S2 indicate that the 
mean baryon fraction within $\delta_c \se 500$ is enhanced by 
at most a factor $1.25$ by dynamical means during  
hierarchical clustering and is more likely 
slightly below the universal value.  This leads to 
two avenues for using the mean intracluster gas fraction 
to constrain $\Omega_m$ which are considered here.  

\subsection{Upper limit approach} 

Because the contribution of baryon sources add linearly, 
the mean baryon content of clusters must 
exceed the mean gas content $\bar{f}_b \ge \fbargas$.  
The mean gas fraction can thus be used to place an 
upper limit on $\Omega_m/\Omega_b$ in eq'n(\ref{fb_rdeltac})
\begin{equation}
\frac{\Omega_m}{\Omega_b} \ \le  \ \Upsilon(\delta_c) \ 
\bar{f}^{-1}_{gas}(\rdeltac) .
\label{ratio_upper_gen}
\end{equation}
From the perspective of maximizing $\Omega_m$, it is appropriate to take 
the $5\%$-ile lower limit on $\fbargas$ 
along with the largest possible baryon enhancement.  
The simulations of White \etal yield an average enhancement 
$\Upsilon(500) \se 1.25$, and it is reasonable to assume this is 
appropriate for the effect on the population mean.  
The upper bound 
on the ratio of total, clustered to baryonic mass density is then 
\begin{equation}
\frac{\Omega_m}{\Omega_b} \ < \ 23.1 \h^{3/2} .
\label{ratio_upper}
\end{equation}
 
Comparison between the light elemental composition of the universe 
and primordial nucleosynthesis expectations places an upper limit on 
the mean baryon density.  This can be combined with the above to express an 
upper limit on $\Omega_m$ directly.  The exact value of the baryon density 
derived in this way 
continues to be a subject of current debate (Krauss \& Kernan 1994; 
Copi, Schramm \& Turner 1995; Steigman 1995; Sasselov \& Goldwirth
1995;  Hata \etal 1996; Tytler, Fan \& Burles 1996; Rugers \& Hogan 1996).  
However, a consensus view is that a firm upper limit 
$\Omega_b \le 0.026 \h^{-2}$ exists simply from the abundanace 
of fragile deuterium in the hostile environment of the local solar
neighborhood (Linsky \etal 1995).  This value produces the constraint
\begin{equation}
\Omega_m \h^{1/2} \ < \ 0.60 .
\label{omega_upper}
\end{equation}
The Einstein--deSitter case $\Omega_m \se 1$ 
requires a very low Hubble constant $h \! \le \! 0.36$ 
(Bartlett \etal 1995).

\subsection{Best estimate approach}

This upper limit is generous in that it : (i) ignores 
the contribution of galaxies to the baryon fraction; (ii) assumes maximal, 
dynamical baryon enhancement through use of an unphysical gas 
equation of state and (iii) uses the largest realistic value of 
$\Omega_b$ combined with the smallest allowed value of 
$\fbargas(\rfiveh)$.  An alternate perspective is to use ``realistic'' 
parameter values to provide a ``best'' estimate of $\Omega_m$.  
For this, I will assume : (i)  
galaxies make a $20 \h^{3/2} \%$ contribution relative to the gas 
$\bar{f}_b(\rfiveh) \se (1+0.2 \h^{3/2}) \fbargas(\rfiveh)$ 
as appropriate for Coma (White \etal 1993) and (ii) a realistic 
equation of state for the gas leads to a modest baryon diminution 
$\Upsilon(500) \se 0.85$.  Utilizing the central value of 
$\fbargas(\rfiveh)$ from eq'n~(\ref{fbargas}) results in 
\begin{equation}
\frac{\Omega_m}{\Omega_b} \ = \ (11.8 \pm 0.7) \ 
\frac{1.2 \h^{3/2}}{1+0.2 \h^{3/2}} \ \simeq \   (11.8 \pm 0.7) \ h^{4/3}
\label{ratio_estimate}
\end{equation}
which is a factor of two smaller than the generous upper limit 
in eq'n~(\ref{ratio_upper}).   The $1\sigma$ error quoted above is 
derived from the propagated statistical error of $\fbargas$.  The
odd--looking slope of $4/3$ in the second expression is a power 
law approximation to the $1.2h^{3/2}/(1+0.2h^{3/2})$ term; it is 
accurate to 2 percent over the range $h \in [0.45, 1]$.  

The final step to estimate $\Omega_m$ requires a value for 
$\Omega_b$.  Recent inferences of the primordial deuterium abundance from 
quasar absorption line spectra produce two different preferred values, 
a low value $\Omega_b h^2 \se 6.2 \pm 0.8 \times 10^{-3}$ (Rugers \& Hogan 
1996) and a high value $\Omega_b h^2 \se 0.024 \pm 0.006$ 
(Tytler, Fan \& Burles 1996)
derived from their respective high and low estimates for D/H.  
These values result in estimates and $1\sigma$ errors of 
\begin{equation}
\Omega_m \h^{2/3} \ = \ 0.07 \pm 0.01 \ \ \ \ \ \
{\rm high~D/H}  ,
\label{highD/H}
\end{equation} 
\begin{equation}
\Omega_m \h^{2/3} \ = \ 0.28 \pm 0.07 \ \ \ \ \ \
{\rm low~D/H} .
\label{lowD/H}
\end{equation}
The latter value should be preferred over the former for several 
reasons.  First, a recent analysis of high resolution Keck spectra of 
Q0014+813 --- the best ``high D/H'' candidate --- 
by Tytler, Burles \& Kirkman (1996) provides no support for a level of
high deuterium absorption, and suggests hydrogen interlopers as a
likely explanation for the data.  In addition, 
the value $\Omega_m \ssim 0.3$ is currently concordant
with large--scale structure observations 
({\it e.g.}, Ostriker \& Steinhardt 1995), 
while the low value $\Omega_m \! < \! 0.1$ deduced from the Rugers \& Hogan
D/H estimate is decidedly difficult in this regard.  

\subsection{Systematic effects}

The modest statistical error in these estimates 
is deceptive, since there are systematic uncertainties in the steps 
involved in the derivation.  In the ``minimal'' approach adopted
above, there is perhaps a factor two uncertainty in the contribution 
of galaxies relative to gas, implying about a $20\%$ error in the 
baryon fraction estimate.  In addition, there is approximately $15\%$ 
uncertainty in the appropriate value of $\Upsilon(500)$.  
Conservatively adding these contributions leads to an estimate of the 
overall systematic uncertainty of $35\%$.  In addition, there
are systematic effects in the nucleosynthesis determination of
$\Omega_b$ which would enlarge the quoted statistical error (Audouze,
Olive \& Truran 1997), but values $\Omega_b h^2 > 0.024$ are unlikely 
because of additional constraints from $^4$He, $^6$Li and $^7$Li 
abundances (\eg, Lemoine \etal 1997).  

The actual error in this analysis could be larger if   
other systematic effects play a significant role.  
Current possibilities include the following.  

\begin{list}
\smallskip
\item {\it (i) Multi--phase intracluster gas} --- One can imagine a 
multi--phase structure, with cooler gas perhaps entrained along loops of 
magnetic field surrounded by hotter, more dilute plasma.  In order 
to be competitive with thermal pressure and create a significant 
mass estimate bias through clumping, 
the magnetic pressure must be close to the thermal 
pressure outside the loops and the mass filling factor within the loops
should be large.  No specific model for the origin and maintenance of 
such a configuration has been proposed.  On energetic grounds alone, 
it is difficult to imagine galaxies supplying enough magnetic field, 
and observations of Faraday rotation in background sources indicate
that any strong fields must have small coherence lengths (Kim \etal
1990; Kronberg 1994). 
Another energetic argument against significant magnetic pressure is 
the fact that the specific thermal energy of the intracluster gas is 
very nearly equal to the specific kinetic energy of the galaxies 
(Jones \& Forman 1984; Lubin \& Bahcall 1993), 
as expected if the thermal and kinetic
pressures respectively support each component within the same
potential well (the assumption underlying the $\beta$-model of 
Cavaliere \& Fusco-Femiano 1976).  
Spatially resolved \xray spectroscopy, particularly of line emission in 
cooler ($2-6 \kev$) clusters, will provide tests of such models, which
are broadly similar to multi--phase cooling flow models (Sarazin 1996).   
Limited information is already available in broad--beam colors 
(Henriksen \& White 1996).  Data from ASCA, SAX, and 
soon XMM and AXAF, coupled with realistic, dynamical and
thermodynamical modeling (Teyssier, Chi\`eze \& Alimi 1996) 
should place stringent constraints on such multiphase models.  

In addition, since the electron pressure structure in the gas 
in a multi-phase model will differ from the standard, 
unclumped model, Sunyaev--Zel'dovich 
measurements can be used to provide independent constraints on clumping.  
Roettiger's (1996) gas dynamic merger simulations show that errors 
in SZ Hubble constant determinations are small in the 
standard scenario, providing hopeful prospects of detecting a modest 
signal from clumpiness.  Present data probably cannot rule out
clumping factors as large as $50\%$, but factors $\gta 2$ 
seem unlikely.  

\smallskip
\item {\it (ii) Mass estimate errors} --- The simulations may be 
giving a misleadingly simple picture of mass estimate accuracy.  The 
good agreement of independent codes employing different gas 
dynamic methods on the same problem 
points the finger at missing physics, rather than numerical inaccuracy, 
if a significant effect is to be found.  Neither feedback from
galactic winds (EMN) nor the inclusion of non-equilibrium 
thermodynamics in the intracluster plasma (Teyssier, Chi\`eze \& Alimi 1996)
significantly alter the virial scaling relation, eq'n~(\ref{M500}).  
The fact that this binding mass estimator is able 
to reconcile the modest WF and DJF sample difference can be taken 
as a measure of empirical support for the simulation results, 
but that could be a misleading interpretation.  
Independent measures of cluster mass, particularly weak gravitational 
lensing (Tyson, Valdes \& Wenk 1990; Kaiser \& Squires 1993) 
are needed to provide additional measures of mass estimate accuracy.
Comparison of the two methods near the virial radius in a few clusters
indicates consistency within modest ($\sim 50\%$) 
statistical error ranges (Squires \etal 1996).  
Extending such studies to larger, statistical samples is
imperative. 

\smallskip
\item {\it (iii) Extrapolation errors} --- 
The \xray data of the WF sample do not extend to $\rfiveh$, making 
extrapolation necessary.  The extrapolation in gas fraction is modest, 
about $1 \h^{-3/2} \%$ on average.  
As a further check on the employed procedure,
the data values at $r_X$ can be used to predict values at 
$0.5 \hinv \mpc$, and the underlying distribution function 
constructed at that radius.  The resultant 5, 50 and 95\% 
confidence limits using the extrapolated gas fraction  
of $9.7$, $13.8$, and $22.9 \%$, respectively (quoted for h=0.5), 
agree well with the directly determined values of $10.0$, 
$13.8$ and $22.3$ from WF (their Figure~6).  Significant errors 
from extrapolation are thus very unlikely.  

\smallskip
\item {\it (iv) Hot dark matter} --- A sea of massive, light neutrinos 
would cluster differently from the cold dark matter typically assumed in the 
dynamical simulations;  their high entropy would prevent them from 
clustering in small potential wells.  Experiments using viable 
cold plus hot dark matter (CHDM) models show that this effect is 
negligible in rich clusters at radii close to $\rfiveh$ (Kofman \etal 1996).  

\smallskip
\item {\it (v) Additional baryon contributions} ---  
Sources of baryons in clusters besides gas and galaxies exist, in the
form of diffuse intracluster stars (Uson, Boughn, \& Kuhn 1991; 
Theuns \& Warren 1996) and MACHOS (Gates, Gyuk, \& Turner 1995; 
Alcock \etal 1996).  
The latter have not been directly detected in intracluster
environments, and they may be simply connected to the former. 
Regardless, extra baryons only drive the constraints on $\Omega_m$ to
lower values.  Their contribution probably does not exceed  
that of galactic starlight.  

\smallskip
\item {\it (vi) Initial baryon fraction inhomogeneities} --- An early 
universe model 
which generates pre-inflationary baryon--to--total mass fluctuations 
and manages to preserve them into the post--inflationary epoch 
would circumvent the causality argument mentioned in \S2.  Such a model 
would need to naturally couple high baryon overdensity to high mass
overdensity in order to produce an overestimate of the cosmic
baryon fraction in clusters.  Such a correlation would also avoid 
producing baryon deficient clusters, a comforting situation since 
no large population of deep, ``empty'' potential wells exists  
(although Bonnet, Mellier \& Fort (1994) have a candidate in the 
field of CL0024+1654).  Consistency
with $\Omega_m \se 1$ would presumably be a natural feature of 
such models.
  
\end{list}

\smallskip
The sources of systematic uncertainty above (except the
last) added in quadrature allow room for perhaps $70\%$ additional
upward error in $\Omega_m$.  However, the magnitude of these 
effects --- particularly of gas clumping for which there are no specific,
dynamical models --- are currently not well understood.  
Forcing all effects in the same
direction could probably manage consistency with $\Omega_m \se
1$ and, in this case, the observational tests cited above should start 
uncovering the effects responsible in the near future.  

\section{Summary}

The gas fraction in clusters of galaxies provides information on 
the cosmic baryon mass fraction.  Dynamical biases of 
clusters' baryon content can be minimized by measuring masses near the 
virial radius, where gas dynamic experiments show equilibrium is 
valid and segregation processes inefficient.  
At $\rfiveh$, the radius where the 
mean interior density is a factor 500 times the critical value, 
the bi--weight mean gas fraction of the 
combined, revised WF and DJF \xray cluster samples is 
$0.060 \pm 0.003 h^{-3/2}$.  This value, when combined with 
our current understanding of cluster formation history and 
limits on $\Omega_b$ from primordial nucleosynthesis, strongly 
favors $\Omega_m \h^{2/3} \sim 0.3$ 
and rules out the possibility of $\Omega_m \se 1$ with high statistical 
significance, unless the Hubble constant is very low $h \! < \! 0.4$.  
These conclusions reinforce, at greater statistical significance, 
earlier work based on different methods and data than those used here (\eg, 
Henriksen \& Mamon 1994; Steigman \& Felten 1995; Lubin \etal 1996).

How serious is the case against $\Omega_m \se 1$?  A die--hard 
Einstein--deSitter advocate could espouse all the elements of the 
analysis leading to the upper limit in eq'n~(\ref{omega_upper}) 
and claim only a small ($30\%$ for $h \se 0.7$) systematic 
effect is missing.  However, 
this approach accounts for neither the hot, \xray emitting 
gas nor the galaxies in clusters, and one might well be suspicious 
of an argument which ignores these two principal, observable components.  
The best estimate approach leaves one short by a factor $\gta 3$.  It is 
possible that this gap is plugged not by a single large effect, but  
by several effects acting in concert, a situation reminiscent 
of the so--called ``$\beta$--discrepancy'' in clusters 
(Evrard 1990b; Lubin \& Bahcall 1995).  
There remains the possibility that a new element of \xray cluster physics 
is simply missing from the present picture.  Observational constraints on 
known sources of systematic error should be vigorously pursued, 
along with more sophisticated theoretical modeling of intracluster
plasma dynamics and thermodynamics.  

The arguments presented here are independent of the value of the 
cosmological constant.  If one favors a spatially flat universe, as 
motivated by simple models of inflation, obtained through a non-zero
cosmological constant $\Lambda$, then the limits on the
clustered matter component predict values of the required 
vacuum energy density $\Omega_\Lambda \equiv 
\Lambda/3\Ho^2 = 1 - \Omega_m$.  For example, for $h=0.7$, the 
upper limit on $\Omega_m$ requires $\Omega_\Lambda \ge 0.28$ while the
best estimate approach gives $\Omega_\Lambda = 0.64 \pm 0.09$.  The 
latter is consistent with the limit $\Omega_\Lambda < 0.66$ derived from 
gravitational lensing arguments (Kochanek 1996) while marginally 
inconsistent with the $\Omega_\Lambda < 0.51$ inferred recently by
Perlemutter \etal (1996) from the magnitude--redshiftt relation for 
Type-Ia supernovae.  

\section*{Acknowledgments}
This work was supported by NASA through Grant NAG5-2790 and by the
CIES and CNRS at the Institut d'Astrophysique in Paris.  This work 
originated at the Institute for Theoretical Physics at UC, 
Santa Barbara, which is supported in part by the National Science 
Foundation under Grant PHY89-04035.  I am grateful to the members and 
staff of the IAP for their warm hospitality during my sabbatical
stay.  I am most grateful to J. Audouze, S. Charlot, G. Mamon, 
J. Mohr, G. Steigman and J. Felten for valuable discussions which 
helped shape the final form of the paper.

\end{document}